\documentclass[aps,prd,twocolumn,10pt,floatfix,superscriptaddress,nofootinbib,secnumarabic]{revtex4-2}

\usepackage{bm}
\usepackage{amsmath}
\usepackage{graphicx}
\usepackage{microtype}
\usepackage{latexsym}
\usepackage{amsfonts}
\usepackage{mathrsfs}
\usepackage{color}
\usepackage{hyperref}

\hypersetup{
	hidelinks,
}

\usepackage[T1]{fontenc}
\usepackage{lmodern,amssymb,mathtools}
\usepackage{xcolor,booktabs,placeins}
\usepackage{bookmark}
\renewcommand{\thesection}{\arabic{section}}
\hypersetup{pdftitle={First subleading correction to the Maxwell dipole greybody factor of a near-extremal Hayward black hole},pdfsubject={Near-extremal Hayward Maxwell dipole scattering}}
\graphicspath{{figure/}}
\AddToHook{env/thebibliography/begin}{\interlinepenalty=10000}
\newcommand{\dd}{\mathrm{d}}
\newcommand{\e}{\mathrm{e}}
\newcommand{\ii}{\mathrm{i}}
\newcommand{\gref}{\Gamma_{\mathrm{ref}}}
\newcommand{\fref}{f_{\mathrm{ref}}}
\newcommand{\bigO}{\mathcal{O}}
\newcommand{\littleo}{o}

\begin{document}

\title{First subleading correction to the Maxwell dipole greybody factor of a near-extremal Hayward black hole}

\author{Hanjun Zou}
\email{zouhanjun24@mails.ucas.ac.cn}
\affiliation{School of Fundamental Physics and Mathematical Sciences, Hangzhou Institute for Advanced Study, UCAS, Hangzhou 310024, China}
\affiliation{School of Physical Sciences, University of Chinese Academy of Sciences, Beijing 100049, China}
\affiliation{Institute of Theoretical Physics, Chinese Academy of Sciences, Beijing 100190, China}

\author{Chenghao Yuan}
\email{yuanchenghao24@mails.ucas.ac.cn}
\affiliation{School of Fundamental Physics and Mathematical Sciences, Hangzhou Institute for Advanced Study, UCAS, Hangzhou 310024, China}
\affiliation{School of Physical Sciences, University of Chinese Academy of Sciences, Beijing 100049, China}
\affiliation{Institute of Theoretical Physics, Chinese Academy of Sciences, Beijing 100190, China}

\author{Zhoujian Cao}
\email[Corresponding author: ]{zjcao@bnu.edu.cn}
\affiliation{School of Fundamental Physics and Mathematical Sciences, Hangzhou Institute for Advanced Study, UCAS, Hangzhou 310024, China}
\affiliation{School of Physics and Astronomy, Beijing Normal University, Beijing 100875, China}

\begin{abstract}
We derive the first finite-gap correction to the Maxwell dipole greybody
factor of a near-extremal Hayward black hole at fixed \(y=\omega/T_H\).
Matching the near-horizon, static, and far-region solutions fixes the
absolute normalization and gives the relative correction
\(d[H(y)+9y/32]\), where \(H(y)\) is defined in
Eq.~\eqref{eq:near-transmission-correction} and \(d\) is the dimensionless
horizon separation.
The two terms arise from near-horizon deformation and long-range Coulomb
normalization, respectively. Integrating the corrected spectrum gives a
dimensionless dipole power per polarization that scales as \((MT_H)^8\),
with a first-order relative increase of approximately \(2.71844\,d\) over
the reference power at the same gap and temperature. At the held-out
temperature \(MT_H\simeq3.84\times10^{-4}\), the correction reduces the
relative power error against the numerical result from \(2.367\%\) to
\(0.083\%\).
\end{abstract}

\maketitle

\allowdisplaybreaks[1]
\raggedbottom

\section{Introduction}

Hawking radiation is a prediction of quantum field theory on curved
spacetime \citep{Hawking1975}. For a static, spherically symmetric black
hole, the spectrum reaching infinity in each bosonic partial wave is
weighted by the thermal occupation at the Hawking temperature \(T_H\)
and by the greybody factor \(\Gamma\), the transmission probability
through the exterior scattering potential \citep{Page1976}.

Regular black-hole models have horizons and nonsingular interiors
\citep{Dymnikova1992,AyonBeatoGarcia1998,FanWang2016}. The Hayward
geometry is a static, asymptotically flat example whose inner and outer
horizons merge at extremality \citep{Hayward2006}. As their dimensionless
separation \(d=(r_{+,\mathrm{phys}}-r_{-,\mathrm{phys}})/M\) tends to zero,
\(MT_H\) tends to zero and a long near-horizon throat develops. The
thermally important frequencies \(\omega=O(T_H)\) probe both the throat
and the exterior scattering potential. We study this regime for the
dipole (\(l=1\)) of an independent massless Maxwell field, its lowest
radiative multipole.

Numerical studies cover greybody factors of axial gravitational
perturbations in a scalar-haired Hayward model \citep{Konoplya2023}
and Hawking emission in the Bardeen spacetime \citep{KonoplyaOvchinnikovAhmedov2023}.
Calz\`a, Pedrotti, and Vagnozzi computed greybody factors and primary
photon spectra for Hayward black holes \citep{CalzaPedrottiVagnozzi2025}.
Direct radial integration, as implemented in GrayHawk
\citep{Calza2025GrayHawk}, and higher-order WKB approximations
\citep{KonoplyaZhidenkoZinhailo2019} provide two ways to compute the
transmission. Comparisons of radial formulations for rapidly rotating
black holes have quantified the sensitivity to the numerical method
\citep{ArbeyCalzaPerezGonzalez2025}. For nearly extremal Hayward black
holes, discrepancies between sixth-order WKB and direct integration
occur at low frequencies and can substantially affect the integrated
Hawking power \citep{CalzaSpina2026}.

Analytic low-frequency absorption calculations \citep{Unruh1976}
include universality results for minimally coupled scalars
\citep{DasGibbonsMathur1997} and matching methods for static black holes
\citep{HarmarkNatarioSchiappa2010}. Near-extremal matching retains the
dependence on frequency relative to the thermal scales
\citep{MaldacenaStrominger1997,CveticLarsen1997}. For Hayward, the static
exterior Maxwell dipole obtained by Yang and Lim supplies the
zero-frequency normalization \citep{YangLim2022}. To calculate the first
correction to the dipole power at small positive \(T_H\), one needs the
absolutely normalized transmission through first order in \(d\) at
fixed \(y=\omega/T_H\), with control of the remainder under the thermal
integral.

We derive the connection directly from the Maxwell radial equation by
matching the near-horizon, static, and far-region solutions. The exact
static dipole fixes the intermediate normalization that connects the
unit-amplitude ingoing wave at the outer horizon to the incident wave
at infinity. The reference spectrum gives the \((M\omega)^4\)
low-frequency law at fixed positive \(d\) and recovers the strictly
extremal \((M\omega)^6\) law along paths with \(y\to\infty\) within the
matching window as \(d\to0^+\). The first relative correction is
\(d[H(y)+9y/32]\). The term \(H(y)\) comes from deforming the
near-horizon wave, and \(9y/32\) comes from the Coulomb normalization
induced by the asymptotic mass term. The Cveti\v{c}--Larsen spin-one formula \citep{CveticLarsen1998}
gives the same first-order frequency dependence after the formal
Hayward substitution in Appendix~\ref{app:cl-hayward}.

We use a uniform remainder estimate on a growing interval in \(y\)
and control the high-frequency tail with \(\Gamma\leq1\) and the
exponential decay of the Bose factor. The dimensionless dipole power
per polarization scales as \((MT_H)^8\) at leading order. Its
first-order increase relative to the reference power at the same
\(d\), evaluated with the exact static coefficient and temperature,
is approximately \(2.71844\,d\). The far-region term supplies about
\(72\%\) of this correction. We compare the spectral and power
predictions with numerical results at four horizon separations,
including a gap reserved before its numerical evaluation.

\mbox{Section}~\ref{sec:geometry} sets up the scattering problem and derives
the reference spectrum and the low-frequency limits.
\mbox{Section}~\ref{sec:spectrum} derives the first near-extremal correction,
and Sec.~\ref{sec:power} calculates the Hawking spectrum and integrated
power. \mbox{Section}~\ref{sec:conclusion} summarizes the results.
Appendix~\ref{app:rn-near} benchmarks the matching on a
Reissner--Nordstr\"om background against known transmission results
\citep{Gubser1997,CveticLarsen1998}. Appendix~\ref{app:numeric} gives
the numerical comparisons.

\section{Geometry, Maxwell equation and normalization}
\label{sec:geometry}

\subsection{Geometry and flux normalization}
We use \(G=c=\hbar=k_B=1\). Let \(M\) denote the mass parameter read from the asymptotically flat metric. The dimensionless coordinates \(r=r_{\mathrm{phys}}/M\) and \(t=t_{\mathrm{phys}}/M\) give the Hayward line element \citep{Hayward2006}
\begin{equation}
 \frac{\dd s^2}{M^2}=-F(r)\dd t^2+\frac{\dd r^2}{F(r)}
       +r^2\dd\Omega^2,
 \label{eq:metric}
\end{equation}
where
\begin{equation}
 F(r)=1-\frac{2r^2}{r^3+a},\qquad a=2(\ell_{\mathrm H}/M)^2.
 \label{eq:lapse}
\end{equation}
In the nonextremal black-hole range the positive horizons are \(r_+\) and \(r_-\). Write \(d=r_+-r_->0\), and factor the lapse as
\begin{equation}
 F=(r-r_+)(r-r_-)g(r;d),
 \label{eq:factorized}
\end{equation}
where
\begin{equation}
 g(r;d)=\frac{r-r_0}{r^3+a},\qquad r_0=2-r_+-r_-.
 \label{eq:factorized-g}
\end{equation}
With frequency \(x=M\omega\), temperature \(\tau=MT_H\), and \(y=x/\tau\), the exact surface gravity relation is
\begin{equation}
 4\pi\tau=F'(r_+)=d\,g_+,\qquad g_+=g(r_+;d).
 \label{eq:temperature}
\end{equation}
At the double root, \(a_e=32/27\), \(r_e=4/3\), and \(g(r_e;0)=1/r_e^2\).

Consider an independent, massless Maxwell field on the fixed background. The separated electromagnetic equations on static spherical geometries are described in Refs.~\citep{CrispinoHiguchiMatsas2001,ArbeyEtAl2021}. Its axial angular components are \(A_A=\e^{-\ii x t}\psi(r)X_A^{lm}\) with a transverse vector harmonic \(X_A^{lm}\) on the unit sphere. Electric-magnetic duality gives the same transmission probability for the other parity. For \(l=1\) the radial equation is
\begin{equation}
 (F\psi')'+\left(\frac{x^2}{F}-\frac{2}{r^2}\right)\psi=0.
 \label{eq:radial}
\end{equation}
The conserved radial flux is
\begin{equation}
 j=\frac{F}{2\ii}\big(\psi^*\psi'-\psi\psi^{*\prime}\big),\qquad
 j'=0.
 \label{eq:radial-flux}
\end{equation}

\begin{figure*}[!t]
 \centering
 \includegraphics[width=\linewidth]{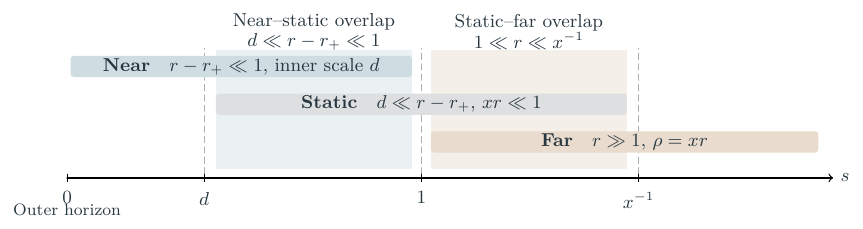}
 \caption{Matching regions for \(0<d\ll1\) and \(0<x\lesssim d\). The shaded intervals mark where the frequency term is small relative to the angular-momentum barrier and adjacent expansions overlap, as in Eqs.~\eqref{eq:near-frequency-ratio} and \eqref{eq:far-frequency-ratio}. The axis shows the ordering of scales in \(s=r-r_+\), with \(s\simeq r\) at large radius.}
 \label{fig:matching-regions}
\end{figure*}

The matching regions in Fig.~\ref{fig:matching-regions} follow from comparing the frequency term \(x^2/F\) with the angular-momentum barrier \(2/r^2\). With \(s=r-r_+\), Eqs.~\eqref{eq:factorized} and \eqref{eq:temperature} give
\begin{equation}
 F=s(s+d)g(r;d),\qquad x=\frac{d g_+y}{4\pi}.
 \label{eq:near-scaling}
\end{equation}
The ratio of the two terms is
\begin{equation}
 \frac{x^2/F}{2/r^2}
 =\frac{x^2r^2}{2s(s+d)g(r;d)}.
 \label{eq:frequency-barrier-ratio}
\end{equation}
Near the outer horizon in the near-extremal regime, Eq.~\eqref{eq:near-scaling} gives
\begin{equation}
 \frac{x^2/F}{2/r^2}
 \sim\frac{y^2}{32\pi^2}\frac{d^2}{s(s+d)}.
 \label{eq:near-frequency-ratio}
\end{equation}
At fixed \(y>0\), the frequency term contributes at leading order on the scale \(s\sim d\). For \(d\ll s\ll1\), the ratio is small and the near-horizon expansion remains valid, giving the near--static overlap.

Far from the black hole, \(F=1+\bigO(r^{-1})\), so
\begin{equation}
 \frac{x^2/F}{2/r^2}
 =\frac{(xr)^2}{2}\bigl[1+\bigO(r^{-1})\bigr].
 \label{eq:far-frequency-ratio}
\end{equation}
Both terms contribute at leading order on the scale \(xr\sim1\). In \(1\ll r\ll x^{-1}\), the far-region expansion remains valid and the frequency term is small relative to the barrier, giving the static--far overlap.

The tortoise coordinate satisfies \(\dd r_*/\dd r=1/F\). We normalize the wave at the outer horizon by \(\psi_{\mathrm H}\sim \e^{-\ii x r_*}\) and write \(\psi_{\mathrm H}\sim A_{\mathrm{in}}\e^{-\ii x r_*}+A_{\mathrm{out}}\e^{\ii x r_*}\) at infinity. Flux conservation gives
\begin{equation}
 |A_{\mathrm{in}}|^2-|A_{\mathrm{out}}|^2=1.
 \label{eq:flux-conservation}
\end{equation}
We define the greybody factor as the ratio of transmitted to incident flux,
\begin{equation}
 \Gamma=|A_{\mathrm{in}}|^{-2},\qquad 0\leq\Gamma\leq 1.
 \label{eq:flux}
\end{equation}

\subsection{Static profile}
We (i) set \(x=0\), so Eq.~\eqref{eq:radial} reduces to \((Fu')'=2u/r^2\), (ii) normalize the growing mode by \(u/r^2\to1/4\) as \(r\to\infty\), and (iii) require regularity at the outer horizon, meaning that \(u\) remains finite as \(r\to r_+^+\). For \(d>0\), these conditions determine the unique solution \(u_d\). We denote its horizon value by \(u_+\).
\begin{equation}
 u_d(r)=\frac{r^2}{4}-\frac{a}{2r},\qquad
 u_+=u_d(r_+) .
 \label{eq:static}
\end{equation}
Direct substitution verifies the static equation, and \(F(r_+)=0\) gives \(a=r_+^2(2-r_+)\) and \(u_+=r_+(3r_+-4)/4>0\). A second solution is \(u_d(r)\int_\infty^r\dd\tilde r/[F(\tilde r)u_d(\tilde r)^2]\). It decays as \(r^{-1}\) at infinity but diverges logarithmically at the simple outer horizon. Regularity excludes this second solution, and the growing-mode normalization fixes the amplitude of \(u_d\).
We use the horizon-normalized static profile
\begin{equation}
 U_{\mathrm{stat}}(r;d)=\frac{u_d(r)}{u_+},\qquad U_{\mathrm{stat}}(r_+;d)=1 .
 \label{eq:static-profile}
\end{equation}

\subsection{Near--static matching}
To match to the static solution at \(x=0\), note that \(x=\tau y=ydg_+/(4\pi)\). The low-frequency limit can thus be reached either by taking \(d\to0^+\) at fixed \(y>0\) or by taking \(y\to0^+\) at fixed \(d>0\). We take the first limit here and introduce the near-horizon variables
\begin{equation}
 z=\frac{r-r_+}{d},\;
 \beta=\frac{x}{d g_+}=\frac{y}{4\pi},\;
 q=\left(\frac{z}{1+z}\right)^{-\ii\beta}.
 \label{eq:nearscales}
\end{equation}
With \(s=dz\), Eq.~\eqref{eq:near-frequency-ratio} tends to \(\beta^2/[2z(1+z)]\) at fixed \(z>0\). The frequency term therefore enters the leading near-region equation. Using Eq.~\eqref{eq:near-scaling}, \(\dd/\dd r=d^{-1}\dd/\dd z\), and \(g(r;d),g_+\to r_e^{-2}\) in Eq.~\eqref{eq:radial}, we obtain
\begin{equation}
 \bigl[z(1+z){\psi_{\mathrm{near}}^{(0)}}'\bigr]'
 +\left(\frac{\beta^2}{z(1+z)}-2\right)\psi_{\mathrm{near}}^{(0)}=0,
 \label{eq:near-leading-equation}
\end{equation}
where primes denote derivatives with respect to \(z\). The ingoing solution is
\begin{equation}
 \psi_{\mathrm{near}}^{(0)}(z)=\left(\frac{z}{1+z}\right)^{-\ii\beta}
   \left(1+\frac{2z}{1-2\ii\beta}\right).
 \label{eq:near-leading-solution}
\end{equation}
In the near--static overlap \(1\ll z\ll d^{-1}\) of Fig.~\ref{fig:matching-regions},
\begin{equation}
 \psi_{\mathrm{near}}^{(0)}(z)\sim\frac{2z}{1-2\ii\beta},\qquad
 U_{\mathrm{stat}}(r_++dz;d)\sim 2z.
 \label{eq:near-static-growth}
\end{equation}
Matching the coefficients of \(z\) gives the static-region wave \(\alpha_0 U_{\mathrm{stat}}\), where
\begin{equation}
 \alpha_0=\frac{1}{1-2\ii\beta}.
 \label{eq:leading-near-connection}
\end{equation}

\subsection{Static--far matching}
In the far region of the same near-extremal limit, hold \(\rho=xr\) fixed as \(x\to0^+\). The leading radial equation is
\begin{equation}
 \frac{\dd^2\psi_{\mathrm{far}}^{(0)}}{\dd\rho^2}
 +\left(1-\frac{2}{\rho^2}\right)\psi_{\mathrm{far}}^{(0)}=0.
 \label{eq:far-leading}
\end{equation}
The solution of this equation that matches the static growth is
\begin{equation}
 \psi_{\mathrm{far}}^{(0)}(\rho)=K\rho j_1(\rho)
 =K\left(\frac{\sin\rho}{\rho}-\cos\rho\right).
 \label{eq:far-leading-solution}
\end{equation}
At small \(\rho\), \(\psi_{\mathrm{far}}^{(0)}\sim K\rho^2/3\). In the static--far overlap \(1\ll r\ll x^{-1}\), the static-region wave grows as \(\alpha_0 U_{\mathrm{stat}}\sim\alpha_0 r^2/(4u_+)\). Matching the coefficients of \(r^2\) gives
\begin{equation}
 \frac{Kx^2}{3}=\frac{\alpha_0}{4u_+},\qquad
 K=\frac{3\alpha_0}{4u_+x^2}.
 \label{eq:far-static-match}
\end{equation}
For \(xr\gg1\), the leading far solution is
\(\psi_{\mathrm{far}}^{(0)}(xr)=-K\cos(xr)+\bigO(|K|/(xr))\).

To obtain the asymptotic wave phases, we set \(w=\sqrt F\,\psi_{\mathrm H}\) in Eq.~\eqref{eq:radial} and retain the terms through order \(1/r\) in the coefficient of \(w\). With \(F=1-2/r+\bigO(r^{-4})\), the large-\(r\) equation at fixed \(x>0\) is
\begin{equation}
 w''+\left[x^2+\frac{4x^2}{r}+\bigO(r^{-2})\right]w=0.
 \label{eq:far-radial-w}
\end{equation}
As \(r\to\infty\), the solution takes the form
\begin{equation}
 w(r)=B_{\mathrm{in}}\e^{-\ii\Phi(r)}
      +B_{\mathrm{out}}\e^{\ii\Phi(r)}+\littleo(1),
 \label{eq:far-waves}
\end{equation}
where
\begin{equation}
 \Phi(r)=xr+2x\log r.
 \label{eq:far-phase}
\end{equation}
With \(r_*=r+2\log r+c_*+\bigO(r^{-1})\), the leading far wave at large \(r_*\) is
\begin{equation}
 \psi_{\mathrm{far}}^{(0)}(r_*)\simeq -K\cos(xr_*-xc_*+\delta_x),
 \label{eq:far-tortoise-wave}
\end{equation}
where \(\delta_x\) is a real constant. Decomposing the cosine into incoming and outgoing waves gives the leading incoming modulus
\begin{equation}
 |A_{\mathrm{in}}|_{\mathrm{ref}}=\frac{|K|}{2}
 =\frac{3}{8u_+x^2|1-2\ii\beta|}.
 \label{eq:ref-incoming}
\end{equation}
Define the exact static coefficient
\begin{equation}
 C_{\mathrm{stat}}(d)=\frac{64}{9}u_+^2
 =\frac49\bigl[r_+(3r_+-4)\bigr]^2 .
 \label{eq:cstat}
\end{equation}
The definition of \(\Gamma\) in Eq.~\eqref{eq:flux} gives the reference spectrum
\begin{equation}
 \gref(d,y)=C_{\mathrm{stat}}(d)x^4\left(1+\frac{y^2}{4\pi^2}\right).
 \label{eq:gamref}
\end{equation}

\subsection{Low-frequency transmission at fixed horizon separation}
In the preceding derivation, we took \(d\to0^+\) at fixed \(y\). We now hold the horizon separation \(d>0\) fixed and take \(y\to0^+\).

Writing \(\psi_{\mathrm H}=\e^{-\ii x r_*}f_x\), we impose \(f_x(r_+)=1\) for a unit-amplitude ingoing wave. The radial equation gives
\begin{equation}
 (Ff_x')'-2\ii x f_x'-\frac{2}{r^2}f_x=0,
 \label{eq:fixed-d-regular-factor}
\end{equation}
where primes denote derivatives with respect to \(r\). Evaluating this equation at the outer horizon and taking \(x\to0^+\) gives
\begin{equation}
 \lim_{x\to0^+} f_x'(r_+)=\frac{2}{r_+^2F'(r_+)}.
\end{equation}
Since \(f_x(r_+)=1\) and the regular static solution satisfies
\begin{equation}
 U_{\mathrm{stat}}(r_+;d)=1,\qquad
 U_{\mathrm{stat}}'(r_+;d)=\frac{2}{r_+^2F'(r_+)},
\end{equation}
\(f_x\) tends to \(U_{\mathrm{stat}}\) at each fixed exterior radius. The static profile thus enters the matching with unit coefficient.

In the overlap \(1\ll r\ll x^{-1}\), matching the static growth \(r^2/(4u_+)\) to the far-region growth \(Kx^2r^2/3\) gives
\begin{equation}
 K=\frac{3}{4u_+x^2},\qquad
 |A_{\mathrm{in}}|\sim\frac{|K|}{2}.
\end{equation}
Using the definition of \(\Gamma\) in Eq.~\eqref{eq:flux}, we obtain
\begin{equation}
 \Gamma\sim C_{\mathrm{stat}}(d)x^4.
 \label{eq:fixed-d-transmission}
\end{equation}
The reference spectrum in Eq.~\eqref{eq:gamref} has the same leading \(x^4\) term as \(y\to0^+\).

\subsection{Low-frequency transmission at extremality}
At \(d=0\), the outer horizon is a double root and \(u_+\) vanishes. We use the unit-amplitude ingoing wave at the double root to fix the normalization of the extremal static solution.
Let \(s=r-r_e\) at \(d=0\). Near the extremal horizon,
\begin{equation}
 F(r)=\frac{s^2}{r_e^2}+\bigO(s^3),\qquad
 r_*\sim-\frac{r_e^2}{s}.
 \label{eq:ext-horizon}
\end{equation}
At \(d=0\), the exact ratio in Eq.~\eqref{eq:frequency-barrier-ratio} and the double-root lapse give
\begin{equation}
 \frac{x^2/F}{2/r^2}\sim\frac{x^2r_e^4}{2s^2}=\frac{\xi^2}{2},
 \qquad \xi=\frac{xr_e^2}{s}.
 \label{eq:ext-frequency-ratio}
\end{equation}
The static approximation requires \(\xi\ll1\), while the double-root expansion requires \(s\ll1\). The strictly extremal near--static overlap is therefore \(x\ll s\ll1\).
Taking \(x,s\to0^+\) at fixed \(\xi\) gives the leading radial equation
\begin{equation}
 \frac{\dd^2\psi_{\mathrm{ext}}^{(0)}}{\dd\xi^2}
 +\left(1-\frac{2}{\xi^2}\right)\psi_{\mathrm{ext}}^{(0)}=0.
 \label{eq:ext-leading-equation}
\end{equation}
The ingoing solution is
\begin{equation}
 \psi_{\mathrm{ext}}^{(0)}(\xi)=\e^{\ii\xi}\left(1+\frac{\ii}{\xi}\right).
 \label{eq:ext-leading-wave}
\end{equation}
In the strictly extremal overlap identified by Eq.~\eqref{eq:ext-frequency-ratio}, the ingoing solution behaves as
\begin{equation}
 \psi_{\mathrm{ext}}^{(0)}\sim\frac{\ii}{\xi}
 =\frac{\ii\,s}{x r_e^2}.
 \label{eq:ext-static-overlap}
\end{equation}
Let \(u_e\) be the static solution in Eq.~\eqref{eq:static} evaluated at extremality. Since \(u_e(r_e)=0\) and \(u_e'(r_e)=1\), it behaves as \(u_e(r_e+s)\sim s\). Writing the leading static-region wave as \(\alpha_{\mathrm{ext}}u_e(r)\), matching the terms linear in \(s\) in Eq.~\eqref{eq:ext-static-overlap} gives
\begin{equation}
 \alpha_{\mathrm{ext}}=\frac{\ii}{x r_e^2}.
 \label{eq:ext-static-amplitude}
\end{equation}
The matched static wave then has the large-\(r\) behavior
\begin{equation}
 \frac{\ii}{x r_e^2}u_e(r)\sim\frac{\ii\,r^2}{4x r_e^2}.
 \label{eq:ext-static-growth}
\end{equation}
Matching this growth to \(K_{\mathrm{ext}}x^2r^2/3\) from Eq.~\eqref{eq:far-leading-solution} in the static--far overlap gives
\begin{equation}
 K_{\mathrm{ext}}=\frac{3\ii}{4r_e^2x^3},\qquad
 |A_{\mathrm{in}}|\sim\frac{|K_{\mathrm{ext}}|}{2}
 =\frac{3}{8r_e^2x^3}.
 \label{eq:ext-far-amplitude}
\end{equation}
It follows that
\begin{equation}
 \Gamma_{\mathrm{ext}}\sim\frac{64}{9}r_e^4x^6.
 \label{eq:extremal}
\end{equation}

\section{The first near-extremal correction}
\label{sec:spectrum}

We derive the first relative correction to the reference spectrum in Eq.~\eqref{eq:gamref} as \(d\to0^+\) at fixed \(y=\omega/T_H>0\). The correction combines the change in the near--static matching amplitude with the change in the far-region normalization.

\subsection{Near-horizon correction}
At fixed \(z=(r-r_+)/d\) and \(\beta=y/(4\pi)\), we retain the terms linear in \(d\) in the near-region radial operator. With \(r=r_++dz\) and \(\widehat g=g(r;d)/g_+\), the exact radial equation is
\begin{equation}
 \bigl[z(1+z)\widehat g\,\psi'\bigr]'
 +\left[\frac{\beta^2}{z(1+z)\widehat g}-\frac{2}{r^2g_+}\right]\psi=0,
 \label{eq:exact-near}
\end{equation}
where primes denote derivatives with respect to \(z\). The coefficients expand as
\begin{equation}
 \begin{aligned}
 \widehat g&=1-dz+\bigO(d^2),\\
 \widehat g^{-1}&=1+dz+\bigO(d^2),\\
 -\frac{2}{r^2g_+}&=-2+d\left(3z+\frac12\right)+\bigO(d^2).
 \end{aligned}
\end{equation}
The radial operator is thus \(\mathcal L_0+d\mathcal L_1+\cdots\), with
\begin{align}
 \mathcal L_0 f
 &=\bigl[z(1+z)f'\bigr]'
   +\left[\frac{\beta^2}{z(1+z)}-2\right]f, \label{eq:leading-near-operator}\\
 \mathcal L_1 f
 &=-\bigl[z^2(1+z)f'\bigr]'
   +\left[\frac{\beta^2}{1+z}+3z+\frac12\right]f.
\end{align}
Using the leading ingoing solution \(\psi_{\mathrm{near}}^{(0)}\) in Eq.~\eqref{eq:near-leading-solution}, we write
\begin{equation}
 \psi=\psi_{\mathrm{near}}^{(0)}
 +d\left(\frac{z}{1+z}\right)^{-\ii\beta}H_1+\cdots.
\end{equation}
We take \(H_1(0)=0\) to preserve the unit ingoing amplitude at the horizon. The coefficient of \(d\) in the radial equation gives
\begin{equation}
 \mathcal L_0\!\left[\left(\frac{z}{1+z}\right)^{-\ii\beta}H_1\right]
 +\mathcal L_1\psi_{\mathrm{near}}^{(0)}=0.
\end{equation}
Define the operator
\begin{equation}
 \mathcal D_\beta f=\bigl[z(1+z)f'\bigr]'-2\ii\beta f'-2f.
\end{equation}
With \(q\) from Eq.~\eqref{eq:nearscales}, \(\mathcal L_0(qH_1)=q\mathcal D_\beta H_1\). Substituting \(\psi_{\mathrm{near}}^{(0)}\) into the first-order equation and cancelling \(q\) gives
\begin{equation}
 \mathcal D_\beta H_1
 =-\frac{1/4+3\beta^2}{\tfrac12-\ii\beta}
  +\frac{2\beta^2(\tfrac12+\ii\beta)}{(\tfrac12-\ii\beta)(1+z)}.
 \label{eq:near-source}
\end{equation}
For real \(\beta\ne0\), a decaying solution of the homogeneous equation \(\mathcal L_0v_\beta=0\) is
\begin{equation}
 v_\beta
 =-\frac{\psi_{\mathrm{near}}^{(0)}}{\tfrac12+\ii\beta}
 -\frac{\overline{\psi_{\mathrm{near}}^{(0)}}-\psi_{\mathrm{near}}^{(0)}}{2\ii\beta}.
\end{equation}
The homogeneous solutions obey
\begin{equation}
 \begin{gathered}
 v_\beta(z)=\bigO(z^{-2})\qquad (z\to\infty),\\
 z(1+z)W[\psi_{\mathrm{near}}^{(0)},v_\beta]=-1,
 \end{gathered}
\end{equation}
where \(W[f,g]=fg'-f'g\) is the Wronskian with respect to \(z\). Variation of parameters gives
\begin{align}
 &(qH_1)(z)
 =\psi_{\mathrm{near}}^{(0)}(z)
    \int_0^z v_\beta(t)q(t)[\mathcal D_\beta H_1](t)\,\dd t \notag\\
 &\qquad-v_\beta(z)
    \int_0^z \psi_{\mathrm{near}}^{(0)}(t)q(t)[\mathcal D_\beta H_1](t)\,\dd t.
 \label{eq:near-variation}
\end{align}
At large \(z\), the first-order correction takes the form
\begin{equation}
 qH_1=h_\infty(\beta)\psi_{\mathrm{near}}^{(0)}+\bigO(1),
\end{equation}
where the relative growth coefficient is
\begin{equation}
 h_\infty(\beta)=\int_0^\infty v_\beta(z)q(z)[\mathcal D_\beta H_1](z)\,\dd z.
\end{equation}
Integration by parts yields
\begin{align}
 h_\infty(\beta)
 &=\frac{\beta^2-\tfrac14}{1-2\ii\beta}
 -\ii\beta\bigl[\Psi(1-2\ii\beta)+\gamma_E\bigr],
 \label{eq:hayward-growth-coefficient}\\
 \operatorname{Re}h_\infty
 &=\frac{2\beta^2}{1+4\beta^2}
 -\frac{\pi\beta}{2}\coth(2\pi\beta),
\end{align}
where \(\Psi\) is the digamma function and \(\gamma_E\) is Euler's constant.
Using this growth coefficient, the near-region wave in the near--static overlap is
\begin{equation}
 \psi_{\mathrm{near}}\sim\frac{2z}{1-2\ii\beta}\bigl[1+dh_\infty\bigr].
 \label{eq:hayward-near-growth}
\end{equation}
For the static profile, \(u_d'(r_+)=1\) and the horizon value expands as
\begin{equation}
 u_+=\frac{r_+^2d g_+}{2}
 =\frac d2\left[1+\frac d4+\bigO(d^2)\right].
 \label{eq:hayward-static-horizon-expansion}
\end{equation}
Consequently, at fixed \(z\),
\begin{equation}
 U_{\mathrm{stat}}(r_++dz;d)
 =1+2z\left(1-\frac d4\right)+\bigO(d^2).
 \label{eq:hayward-static-overlap-expansion}
\end{equation}
Taking the ratio of the linear growth terms at fixed \(\beta\) gives the matching amplitude through first order,
\begin{equation}
 \begin{aligned}
 \alpha
 &=\frac{2z(1+dh_\infty)/(1-2\ii\beta)}{2z(1-d/4)}+\bigO(d^2)\\
 &=\frac{1}{1-2\ii\beta}
 \left[1+d\left(h_\infty+\frac14\right)\right]+\bigO(d^2).
 \end{aligned}
 \label{eq:H-alpha}
\end{equation}
Since \(\Gamma=|A_{\mathrm{in}}|^{-2}\), the near-region transmission factor relative to the leading amplitude \(\alpha_0\) in Eq.~\eqref{eq:leading-near-connection} is
\begin{equation}
 \begin{aligned}
 \frac{|\alpha_0|^2}{|\alpha|^2}
 &=1-2d\,\operatorname{Re}\!\left(h_\infty+\frac14\right)+\bigO(d^2)\\
 &=1+dH(y)+\bigO(d^2),
 \label{eq:hayward-near-transmission-factor}
 \end{aligned}
\end{equation}
where
\begin{equation}
 \begin{aligned}
 H(y)&=-2\operatorname{Re}\!\left(h_\infty+\frac14\right)\\
 &=\frac{y}{4}\coth\frac y2-\frac12-\frac{y^2}{4\pi^2+y^2}.
 \end{aligned}
 \label{eq:near-transmission-correction}
\end{equation}
As \(y\to0\),
\begin{equation}
 H(y)=\left(\frac1{24}-\frac1{4\pi^2}\right)y^2+\bigO(y^4).
\end{equation}

\subsection{Far-region normalization}
The near--static matching gives the wave \(\alpha U_{\mathrm{stat}}\) in the static region. With \(\rho=xr\), its growing term in the static--far overlap \(1\ll r\ll x^{-1}\) is
\begin{equation}
 \psi\sim\alpha U_{\mathrm{stat}}
 \sim\frac{\alpha r^2}{4u_+}
 =\frac{\alpha}{4u_+x^2}\rho^2.
 \label{eq:static-far-growth}
\end{equation}
To eliminate the first-derivative term in the far-region equation, set \(w=\sqrt F\,\psi\). Expanding at fixed \(\rho>0\) with \(F(\rho/x)=1-2x/\rho+\bigO_\rho(x^4)\) gives
\begin{equation}
 \frac{\dd^2w}{\dd\rho^2}
 +\left[1-\frac{2}{\rho^2}+\frac{4x}{\rho}
 -\frac{2x}{\rho^3}+\bigO_\rho(x^2)\right]w=0.
 \label{eq:far-potential-body}
\end{equation}
The growing solution of Eq.~\eqref{eq:far-potential-body} through first order in \(x\) is
\begin{equation}
 \frac{w(\rho)}{\mathcal N}\simeq
 \frac{F_1(-2x,\rho)}{C_1(-2x)}+3x(-f_1J+g_1I),
 \label{eq:far-first-order-solution}
\end{equation}
where \(\mathcal N\) is fixed by matching to the static solution. Here \(F_1(\eta,\rho)\) is the regular Coulomb wave for \(\ell=1\) \citep{DLMF}, with \(\eta=-2x\) set by the \(4x/\rho\) term. Its small-\(\rho\) normalization coefficient is
\begin{equation}
 C_1(\eta)=\frac{2\e^{-\pi\eta/2}|\Gamma_{\mathrm E}(2+\ii\eta)|}{3!},
\end{equation}
where \(\Gamma_{\mathrm E}\) is the Euler Gamma function.
The functions \(f_1\) and \(g_1\) in Eq.~\eqref{eq:far-first-order-solution} are
\begin{equation}
 \begin{aligned}
 f_1(\rho)&=\rho j_1(\rho)=\frac{\sin\rho}{\rho}-\cos\rho,\\
 g_1(\rho)&=\rho y_1(\rho)=-\frac{\cos\rho}{\rho}-\sin\rho.
 \end{aligned}
 \label{eq:far-spherical-basis}
\end{equation}
The functions \(J\) and \(I\) are
\begin{equation}
 \begin{aligned}
 J(\rho)&=-\frac{\cos(2\rho)}{2\rho^3}+\frac{\sin(2\rho)}{4\rho^4},\\
 I(\rho)&=\frac12\left[1-\frac{1}{\rho^2}
 +\frac{\sin(2\rho)}{\rho^3}-\frac{\sin^2\rho}{\rho^4}\right].
 \end{aligned}
 \label{eq:far-variation-functions}
\end{equation}
In the static--far overlap \(x\ll\rho\ll1\), the two terms in the far-region solution have the expansions
\begin{equation}
 \begin{aligned}
 \frac{F_1(-2x,\rho)}{C_1(-2x)}
 &=\rho^2\bigl[1-x\rho+\bigO(\rho^2)\bigr],\\
 3x(-f_1J+g_1I)&=-x\rho+\bigO(x\rho^3).
 \end{aligned}
 \label{eq:far-small-rho}
\end{equation}
On the static side, Eq.~\eqref{eq:static-far-growth} and \(\sqrt F=1-1/r+\bigO(r^{-2})\) give
\begin{equation}
 \alpha\sqrt F\,U_{\mathrm{stat}}
 =\frac{\alpha}{4u_+x^2}\bigl[\rho^2-x\rho+\bigO(x^2)\bigr].
 \label{eq:matched-static-small-rho}
\end{equation}
Matching the coefficients of \(\rho^2\) fixes \(\mathcal N=\alpha/(4u_+x^2)\). The same normalization gives agreement of the \(-x\rho\) terms.

At infinity,
\begin{equation}
 J(\infty)=0,\qquad
 I(\infty)=\int_0^\infty\frac{2f_1(t)^2}{t^3}\,\dd t=\frac12.
\end{equation}
The matched wave at large \(r_*\) therefore takes the form
\begin{equation}
 w\simeq\frac{\alpha}{4u_+x^2}
 \left[\frac{\sin\theta}{C_1(-2x)}-\frac{3x}{2}\cos\theta\right],
 \label{eq:far-matched-phase}
\end{equation}
where \(\theta=xr_*+\delta\) and \(\delta\) is a real phase independent of \(r_*\). Expanding the sine and cosine into traveling waves gives
\begin{equation}
 w\simeq\mathcal N\left[
 \left(\frac{\ii}{2C_1(-2x)}-\frac{3x}{4}\right)
 \e^{-\ii\delta}\e^{-\ii xr_*}+\mathrm{h.c.}
 \right].
 \label{eq:far-matched-traveling-waves}
\end{equation}
\subsection{Combined correction and uniform remainder}
The coefficient of \(\e^{-\ii xr_*}\) gives the incoming amplitude,
\begin{equation}
 \begin{aligned}
 A_{\mathrm{in}}
 &=\frac{\alpha}{4u_+x^2}\e^{-\ii\delta}
 \Biggl[\frac{\ii}{2C_1(-2x)}-\frac{3x}{4}\\
 &\qquad+\bigO\!\left(d^{4/3}P(1+y)+x^{4/3}\right)\Biggr].
 \end{aligned}
 \label{eq:matched-incoming}
\end{equation}
The fixed polynomial \(P\) bounds the \(y\)-dependence of the \(d^{4/3}\) remainder. Taking the inverse squared modulus gives
\begin{equation}
 \frac{\Gamma}{\gref}
 =\left|\frac{\alpha_0 C_1(-2x)}{\alpha C_1(0)}\right|^2
 +\bigO\bigl(d^{4/3}P(1+y)+x^{4/3}\bigr).
 \label{eq:normalized-transmission-factors}
\end{equation}
The Coulomb factor contributes \(2\pi x\) at first order. Using \(2\pi x=(9y/32)d+\bigO(d^2y)\), we obtain
\begin{equation}
 \frac{\Gamma}{\gref}
 =1+d B_{\mathrm H}(y)+\bigO\!\left(d^{4/3}P(1+y)\right),
 \label{eq:central}
\end{equation}
where
\begin{equation}
 B_{\mathrm H}(y)=\frac{9y}{32}+H(y).
 \label{eq:hayward-first-order-coefficient}
\end{equation}
It can be verified that, for a fixed polynomial \(P\), the remainder estimate in Eq.~\eqref{eq:central} holds uniformly as \(d\to0^+\) in the window
\begin{equation}
 0<y\le Y(d)=9\log(1/d).
\end{equation}
Along paths in this window with \(y\to\infty\) as \(d\to0^+\), \(\Gamma/\gref\to1\) and \(\gref/x^6\to(64/9)r_e^4\), recovering the \(x^6\) law and its coefficient in Eq.~\eqref{eq:extremal}.

\section{Hawking spectrum and integrated power}
\label{sec:power}

\begin{figure*}[!t]
 \centering
 \includegraphics[width=\linewidth]{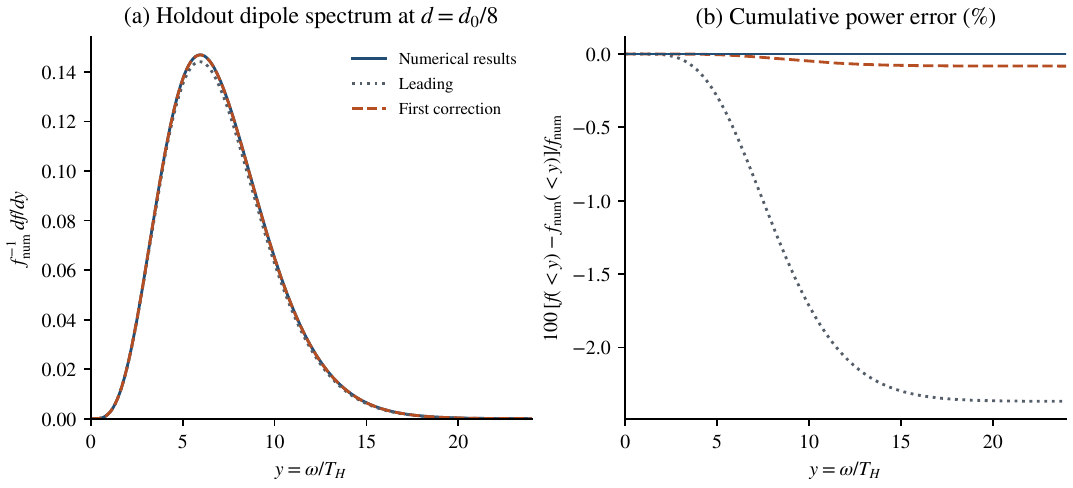}
 \caption{Dipole spectrum at \(\tau=3.83577\times10^{-4}\) for the held-out gap \(d=d_0/8\), with \(d_0=0.0688486\). (a) The reference spectrum, the first-order prediction, and the numerical results, all normalized by the numerical power \(f_{\mathrm{num}}\). (b) Cumulative power errors relative to \(f_{\mathrm{num}}\).}
 \label{fig:thermal}
\end{figure*}

Define \(f_{1,1}=M^2 P_{1,1}\) as the power in the \(l=1\) Maxwell mode for one polarization, summed over \(m=-1,0,1\). With the Hawking occupation factor, its density is \citep{Hawking1975,Page1976}
\begin{equation}
 \frac{\dd f_{1,1}}{\dd y}
  =\frac{3\tau^2}{2\pi}\frac{y}{\e^y-1}\,\Gamma(d,y).
 \label{eq:power-density}
\end{equation}
The two polarizations together give twice the power in Eq.~\eqref{eq:power-density}. Inserting \(\gref\) at the same \(d\) and \(\tau\) gives the integrated reference power for one polarization,
\begin{equation}
 \fref=\frac{3}{2\pi}C_{\mathrm{stat}}(d)\tau^6 Q,
 \label{eq:reference-power}
\end{equation}
where
\begin{equation}
 \begin{aligned}
 Q&=\int_0^\infty w_T(y)\,\dd y,\\
 w_T(y)&=\frac{y^5\left(1+y^2/(4\pi^2)\right)}{\e^y-1}.
 \end{aligned}
 \label{eq:thermal-moments}
\end{equation}
As \(d\to0^+\), the reference power scales as
\begin{equation}
 \fref\sim\frac{8Q}{3\pi(4\pi r_e^2)^6}\,d^8.
\end{equation}

The power density in Eq.~\eqref{eq:power-density} and the reference weight in Eq.~\eqref{eq:thermal-moments} give
\begin{equation}
 \frac{f_{1,1}}{\fref}
 =\frac1Q\int_0^\infty w_T(y)\frac{\Gamma(d,y)}{\gref(d,y)}\,\dd y.
\end{equation}
On \(0<y\le Y(d)\), the uniform estimate in Eq.~\eqref{eq:central} bounds the integrated relative remainder by a constant times
\begin{equation}
 \frac{d^{4/3}}{Q}\int_0^\infty w_T(y)P(1+y)\,\dd y
 =\bigO(d^{4/3}).
\end{equation}
The integral is finite because \(w_T(y)\) behaves as \(y^4\) at zero and decays exponentially at infinity.

For \(y\ge Y(d)\), Eq.~\eqref{eq:flux} gives \(\Gamma\le1\), and \((\e^y-1)^{-1}\le \e^{-y}/(1-\e^{-Y})\). With \(Y=9\log(1/d)\) and \(\tau^2/\fref=\bigO(d^{-6})\) as \(d\to0^+\), the tail satisfies
\begin{align}
 &\frac{1}{\fref}\int_Y^\infty\frac{\dd f_{1,1}}{\dd y}\,\dd y
 \leq \frac{3\tau^2}{2\pi\fref}\int_Y^\infty\frac{y}{\e^y-1}\,\dd y \notag\\
 &\qquad\leq \frac{3\tau^2}{2\pi\fref}\frac{Y+1}{\e^Y-1}
 =\bigO(d^3|\log d|).
 \label{eq:tail-bound}
\end{align}
The tails of \(w_T\) and \(w_T B_{\mathrm H}\) also decay exponentially. Thus the first-order power correction involves the moment of \(B_{\mathrm H}\) over the full frequency range,
\begin{equation}
 \frac{f_{1,1}}{\fref}
 =1+\frac{d}{Q}\int_0^\infty w_T(y)B_{\mathrm H}(y)\,\dd y+\littleo(d).
\end{equation}
Define the first moment
\begin{equation}
 Q_y=\int_0^\infty y\,w_T(y)\,\dd y .
\end{equation}
Evaluating the integral of \(w_T B_{\mathrm H}\) gives
\begin{equation}
 \frac{f_{1,1}}{\fref}
 =1+\left(\frac52+\frac{Q_y}{32Q}\right)d+\littleo(d).
 \label{eq:power-ref}
\end{equation}
The linear coefficient is \(\frac52+Q_y/(32Q)\approx2.71844\).
Using \(u_+/\tau=2\pi r_+^2\) and \(r_+=r_e+d/2+\bigO(d^2)\), we obtain
\begin{equation}
 \frac{C_{\mathrm{stat}}(d)}{\tau^2}
 =\frac{256\pi^2}{9}r_e^4\left[1+\frac32d+\bigO(d^2)\right].
\end{equation}
Combining Eqs.~\eqref{eq:reference-power} and \eqref{eq:power-ref} then gives
\begin{align}
 \frac{f_{1,1}}{\tau^8}
 &=\frac{f_{1,1}}{\fref}\frac{\fref}{\tau^8}
 =\frac{3Q}{2\pi}\frac{C_{\mathrm{stat}}(d)}{\tau^2}
   \frac{f_{1,1}}{\fref} \notag\\
 &=\frac{128\pi Q r_e^4}{3}
   \left[1+\left(4+\frac{Q_y}{32Q}\right)d+\littleo(d)\right].
 \label{eq:power-tau}
\end{align}
Numerically, \(128\pi Qr_e^4/3\approx1.06022\times10^5\) and \(4+Q_y/(32Q)\approx4.21844\).
In the normalization of Eq.~\eqref{eq:power-ref}, the far- and near-region coefficients are \(9Q_y/(32Q)\approx1.96596\) and \(5/2-Q_y/(4Q)\approx0.752482\), respectively. The long-range mass term supplies approximately \(72.3193\%\) of this linear correction.

Figure~\ref{fig:thermal} compares the spectrum corrected by \(dB_{\mathrm H}(y)\) in Eq.~\eqref{eq:central} with the reference spectrum and the numerical results. At the held-out temperature, the correction in Eq.~\eqref{eq:power-ref} reduces the relative error in the integrated power from \(2.367\%\) to \(0.083\%\). Appendix~\ref{app:numeric} gives the comparisons across four horizon gaps.

\FloatBarrier

\section{Conclusion}
\label{sec:conclusion}

At fixed \(y=\omega/T_H\), the first relative correction to the
reference Maxwell dipole greybody factor of a near-extremal Hayward
black hole is \(dB_{\mathrm H}(y)=d[H(y)+9y/32]\) (see Eq.~\eqref{eq:central}), where \(d=r_+-r_-\) is the horizon separation
in units of the mass. The near-horizon deformation gives \(H(y)\),
and the long-range Coulomb normalization set by the asymptotic mass
term gives \(9y/32\). The reference spectrum reproduces the
\((M\omega)^4\) low-frequency law at fixed \(d>0\). It recovers the
strictly extremal \((M\omega)^6\) law and its coefficient along paths
with \(y\to\infty\) inside the matching window as \(d\to0^+\).

The uniform remainder estimate on \(0<y\leq9\log(1/d)\) and the
exponential suppression of the high-frequency tail control the
thermal integral through first order. The dimensionless power
\(f_{1,1}\) for one polarization, summed over the three dipole states,
scales as \(\tau^8\) at leading order, with \(\tau=MT_H\).
The first-order increase relative to the reference power evaluated
with the exact static coefficient and temperature at the same \(d\)
is approximately \(2.71844\,d\) (Eq.~\eqref{eq:power-ref}). The
far-region term supplies about \(72\%\) of this increase. The relative
coefficient of \(d\) in \(f_{1,1}/\tau^8\) is approximately \(4.21844\)
(Eq.~\eqref{eq:power-tau}), including the variation of
\(C_{\mathrm{stat}}/\tau^2\) with \(d\).

The Reissner--Nordstr\"om benchmark recovers known leading and
first-order transmission results (Appendix~\ref{app:rn-near}). For Hayward,
numerical calculations at four horizon separations show that the scaled
correction approaches \(B_{\mathrm H}(y)\) at the sampled frequencies
as \(d\) decreases. At the held-out temperature \(MT_H\simeq3.84\times10^{-4}\),
the first-order correction reduces the relative error in the integrated
power from \(2.367\%\) to \(0.083\%\) (Appendix~\ref{app:numeric},
Table~\ref{tab:power}).

\begin{acknowledgments}

We thank Meng Xu for carefully reviewing the manuscript and helping to revise it.

This work is supported by the National Key Research and Development Program of China (No.~2021YFC2203001) and the National Natural Science Foundation of China (No.~12475049).

\end{acknowledgments}

\section*{Statements and Declarations}

\subsection*{Data availability}

The numerical data underlying the figures and tables are available from the corresponding author upon reasonable request.

\FloatBarrier
\appendix
\renewcommand{\thesubsection}{\thesection.\arabic{subsection}}
\section{Reissner--Nordstr\"om and Cveti\v{c}--Larsen comparisons}
\label{app:rn}

\subsection{Near-horizon correction in Hayward and Reissner--Nordstr\"om}
\label{app:rn-near}

The Hayward and Reissner--Nordstr\"om (RN) geometries both have
\(L_2/r_e=1\) at extremality and share the leading equation for an
independent Maxwell dipole in the throat. Their first-order corrections
depend on the cubic term in the extremal lapse, measured by
\begin{equation}
 k=\frac{F'''(r_e)}{3F''(r_e)}.
 \label{eq:cubic}
\end{equation}
In the mass-normalized radius, \((r_e,k)=(4/3,-1)\) for Hayward and
\((1,-2)\) for RN.
The growing terms of the RN near-region solution, corresponding to
Eq.~\eqref{eq:hayward-near-growth}, are
\begin{equation}
 \psi_{\mathrm{near}}^{\mathrm{RN}}
 \simeq 2z\alpha_0
 \left[1+d\left(\frac z2+\widetilde h_\infty^{\mathrm{RN}}
 +\frac{\ii\beta}{2}\right)\right],
\end{equation}
where \(\alpha_0=(1-2\ii\beta)^{-1}\) and
\(\widetilde h_\infty^{\mathrm{RN}}=2[h_\infty(\beta)+1/4]\), with
\(h_\infty\) from Eq.~\eqref{eq:hayward-growth-coefficient}.
The exact RN static solution is \(u_d^{\mathrm{RN}}=(r^2-r_+r_-)/4\),
with \(r_\pm=1\pm d/2\). Its horizon value, the counterpart of
Eq.~\eqref{eq:hayward-static-horizon-expansion}, is
\begin{equation}
 u_+^{\mathrm{RN}}=\frac{r_+d}{4}.
\end{equation}
This gives \(C_{\mathrm{stat}}^{\mathrm{RN}}=(4/9)(r_+d)^2\).
The horizon-normalized static profile corresponding to
Eq.~\eqref{eq:hayward-static-overlap-expansion} is
\begin{equation}
 U_{\mathrm{stat}}^{\mathrm{RN}}(r_++dz;d)=1+2z+\frac{d}{r_+}z^2.
\end{equation}
Taking the ratio of the growing terms as in Eq.~\eqref{eq:H-alpha}
gives the matching amplitude through first order,
\begin{equation}
 \begin{aligned}
 \alpha_{\mathrm{RN}}
 &\simeq\alpha_0
 \frac{1+d(z/2+\widetilde h_\infty^{\mathrm{RN}}+\ii\beta/2)}
 {1+dz/2}\\
 &\simeq\alpha_0\left[1+d\left(
 \widetilde h_\infty^{\mathrm{RN}}+\frac{\ii\beta}{2}\right)\right].
 \end{aligned}
\end{equation}
The near-region transmission factor for RN, corresponding to
Eq.~\eqref{eq:hayward-near-transmission-factor}, is
\begin{equation}
 \begin{aligned}
 \frac{|\alpha_0|^2}{|\alpha_{\mathrm{RN}}|^2}
 &=1-2d\operatorname{Re}\!\left(
 \widetilde h_\infty^{\mathrm{RN}}+\frac{\ii\beta}{2}\right)
 +\bigO(d^2)\\
 &=1+2dH(y)+\bigO(d^2).
 \end{aligned}
 \label{eq:rn-near-transmission-factor}
\end{equation}
The term \(\ii\beta/2\) contributes only to the phase at this order.

As in Eq.~\eqref{eq:normalized-transmission-factors}, combining the
near-region matching factor with the far-region Coulomb factor gives,
for either background at fixed \(y\),
\begin{equation}
 \frac{\Gamma}{\gref}
 =\frac{|\alpha_0|^2}{|\alpha|^2}
 \left|\frac{C_1(-2x)}{C_1(0)}\right|^2+\littleo(d),
 \label{eq:rn-transmission-factors}
\end{equation}
where \(\alpha\) is the matching amplitude of the corresponding
background. Using the near-region factors in
Eqs.~\eqref{eq:hayward-near-transmission-factor} and
\eqref{eq:rn-near-transmission-factor} and the Coulomb factor
\(1+2\pi x+\bigO(x^2)\) gives
\begin{equation}
 \frac{\Gamma}{\gref}=1-kdH(y)+2\pi x+\littleo(d).
 \label{eq:rn-hayward-relative-correction}
\end{equation}
For RN, \(k=-2\) and \(x=\tau y=dy/(4\pi)+\bigO(d^2)\) at fixed
\(y\). The counterparts of Eqs.~\eqref{eq:central} and
\eqref{eq:hayward-first-order-coefficient} are therefore
\begin{equation}
 \begin{aligned}
 \frac{\Gamma_{\mathrm{RN}}}{\Gamma_{\mathrm{ref}}^{\mathrm{RN}}}
 &=1+dB_{\mathrm{RN}}(y)+\littleo(d),\\
 B_{\mathrm{RN}}(y)&=2H(y)+\frac{y}{2}.
 \end{aligned}
 \label{eq:rn-first-order-coefficient}
\end{equation}
To compare the throat corrections, we subtract \(2\pi x\) using
\(x=\tau y\) for each background. At fixed \(y\), using
\(d/\tau\to4\pi r_e^2\) gives
\begin{equation}
 \lim_{\tau\to0}\frac{\Gamma/\gref-1-2\pi x}{\tau}
 =-4\pi r_e^2k\,H(y).
 \label{eq:rn-comparison}
\end{equation}
The leading RN probability agrees with Gubser's dipole result
\citep{Gubser1997}. Its first-order relative correction
\(2\pi x+2dH(y)\) agrees with the Cveti\v{c}--Larsen expansion
\citep{CveticLarsen1998}.

\begin{figure}[!htbp]
 \centering
 \includegraphics[width=\linewidth]{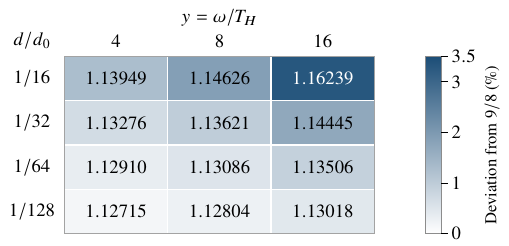}
 \caption{Convergence of the near-horizon correction in the RN and
 Hayward backgrounds. The cells show \(S_{\mathrm{RN}}/S_{\mathrm H}\)
 at three fixed frequencies and four horizon separations. Color gives
 the absolute relative deviation from the analytic limit \(9/8=1.125\),
 expressed as a percentage.}
 \label{fig:rn-convergence}
\end{figure}

For the numerical comparison in Fig.~\ref{fig:rn-convergence}, define
\(S(d,y)=(\Gamma/\gref-1-2\pi x)/\tau\) using the exact static
normalization and temperature of each background. At the same \(d\)
and \(y\), the ratio \(S_{\mathrm{RN}}/S_{\mathrm H}\) approaches the
value \(9/8\) predicted by Eq.~\eqref{eq:rn-comparison}.

\FloatBarrier

\subsection{Hayward comparison with the Cveti\v{c}--Larsen frequency factors}
\label{app:cl-hayward}

The coefficient \(k\) in Eq.~\eqref{eq:cubic} also enters the
Cveti\v{c}--Larsen frequency factors through the two horizon surface
gravities. We take \(j=s=1\) in their absorption formula,
Eq.~(97) of Ref.~\citep{CveticLarsen1998}, and formally substitute the Hayward
surface gravities. With \(g_\pm=g(r_\pm;d)\), their positive magnitudes
are \(\kappa_\pm=dg_\pm/(2M)\). The inverse-temperature combinations
in Eqs.~(18)--(20) of Ref.~\citep{CveticLarsen1998}
give the Gamma-function arguments
\begin{align}
 a_{\mathrm H}&=\frac{y}{2\pi},\qquad
 a_{\mathrm R}=a_{\mathrm H}-a_{\mathrm L},\notag\\
 a_{\mathrm L}&=\frac{\omega}{2}
  \left(\frac{1}{\kappa_+}-\frac{1}{\kappa_-}\right).
\end{align}
Since \(g_+/g_-=1+kd+\bigO(d^2)\), at fixed \(y\) we obtain
\begin{equation}
 a_{\mathrm L}
 =\frac{a_{\mathrm H}}{2}\left(1-\frac{g_+}{g_-}\right)
 =-\frac{kd}{2}a_{\mathrm H}+\bigO(d^2).
\end{equation}
For the spin-one dipole, the horizon-dependent factor is
\begin{equation}
 \mathcal T=
 \frac{|\Gamma_{\mathrm E}(2+\ii a_{\mathrm R})|^2
       |\Gamma_{\mathrm E}(1+\ii a_{\mathrm L})|^2}
      {|\Gamma_{\mathrm E}(1+\ii a_{\mathrm H})|^2}.
\end{equation}
At \(a_{\mathrm L}=0\), it reduces to \(1+a_{\mathrm H}^2\), the
thermal factor in Eq.~\eqref{eq:gamref}. Expanding at fixed
\(a_{\mathrm H}\) gives
\begin{equation}
 \frac{\mathcal T}{1+a_{\mathrm H}^2}
 =1-a_{\mathrm L}\frac{\dd}{\dd a_{\mathrm H}}
 \log|\Gamma_{\mathrm E}(2+\ii a_{\mathrm H})|^2+\bigO(d^2).
\end{equation}
Using \(|\Gamma_{\mathrm E}(2+\ii a)|^2=(1+a^2)\pi a/\sinh(\pi a)\)
and \(H(y)\) from Eq.~\eqref{eq:near-transmission-correction}, we find
\begin{equation}
 \begin{gathered}
 \frac{a_{\mathrm H}}{2}\frac{\dd}{\dd a_{\mathrm H}}
 \log|\Gamma_{\mathrm E}(2+\ii a_{\mathrm H})|^2
 =-H(y),\\
 \frac{\mathcal T}{1+a_{\mathrm H}^2}
 =1-kdH(y)+\bigO(d^2).
 \end{gathered}
\end{equation}

In the mass convention of Cveti\v{c} and Larsen, \(G_{\mathrm N}=1/8\),
so \(M_{\mathrm{CL}}\omega=8x\) in our notation. Dividing their
Coulomb factor by its zero-frequency value gives
\begin{equation}
 \mathcal C(x)=\e^{2\pi x}
 \frac{|\Gamma_{\mathrm E}(3+2\ii x)|^2}{|\Gamma_{\mathrm E}(3)|^2}
 =1+2\pi x+\bigO(x^2).
\end{equation}
We fix the coefficient of \(x^4\) in the fixed-gap low-frequency
limit using the exact Hayward static coefficient
\(C_{\mathrm{stat}}(d)\). Relative to \(\gref\), the resulting
frequency factor at fixed \(y\) is
\begin{equation}
 \frac{\mathcal C(x)\mathcal T}{1+a_{\mathrm H}^2}
 =1+d\left[H(y)+\frac{9y}{32}\right]+\bigO(d^2),
\end{equation}
where we used \(k=-1\) and \(2\pi x=(9y/32)d+\bigO(d^2)\).
This reproduces the first-order frequency dependence obtained by
direct radial matching in Eq.~\eqref{eq:central}.

\onecolumngrid
\section{Numerical comparisons}
\label{app:numeric}

The four gaps are \(d=d_0,d_0/2,d_0/4,d_0/8\), with \(d_0=0.0688486\) corresponding to \(\ell_{\mathrm H}/\ell_{\mathrm{ext}}=0.999\) and \(\ell_{\mathrm{ext}}/M=4/(3\sqrt{3})\). Table~\ref{tab:power} compares the reference power \(\fref\), the first-order prediction \(f_{\mathrm{pred}}\), and the numerical power \(f_{\mathrm{num}}\). Its caption lists the numerical integration windows.

For each horizon separation, we choose the upper endpoint \(y_{\max}\) so that the high-frequency tail bound is at most \(2\times10^{-9}\fref\). In power units, Eq.~\eqref{eq:tail-bound} gives
\[
 f_{\mathrm{tail}}(Y)
 =\int_Y^\infty\frac{\dd f_{1,1}}{\dd y}\,\dd y
 \le \frac{3\tau^2}{2\pi}\frac{Y+1}{\e^Y-1},
\]
with \(Y=y_{\max}\) for each window. Below the common lower endpoint \(y_{\min}=1/64\), we use the static approximation in Eq.~\eqref{eq:fixed-d-transmission}. The resulting low-frequency contribution is
\[
 f_{\mathrm{low}}^{\mathrm{stat}}
 =\frac{3C_{\mathrm{stat}}(d)\tau^6}{2\pi}
 \int_0^{y_{\min}}\frac{y^5}{\e^y-1}\,\dd y.
\]
This contribution is less than \(7.3\times10^{-13}f_{\mathrm{num}}\) for all four horizon separations.

\begin{table}[!hb]
 \caption{Dimensionless dipole power for one polarization and three \(m\) states.
 The signed errors are \(e_0=(\fref-f_{\mathrm{num}})/f_{\mathrm{num}}\) and \(e_1=(f_{\mathrm{pred}}-f_{\mathrm{num}})/f_{\mathrm{num}}\).
 The absolute spectral difference \(\Delta_{\mathrm{abs}}\) is the integral of \(\lvert \dd f_{\mathrm{pred}}/\dd y-\dd f_{\mathrm{num}}/\dd y\rvert\) from \(1/64\) to \(y_{\max}=47,51,55,59\), respectively.}
 \label{tab:power}
 \centering
 \small
 \setlength{\tabcolsep}{2.4pt}
 \begin{tabular}{@{}lrrrrrrr@{}}
 \toprule
 \(d/d_0\) & \(\tau\) & \(\fref\) & \(f_{\mathrm{pred}}\)
 & \(f_{\mathrm{num}}\) & \(\Delta_{\mathrm{abs}}\)
 & \(e_0\) (\%) & \(e_1\) (\%) \\
 \midrule
 \(1\) & \(2.97904\,10^{-3}\) & \(7.27637\,10^{-16}\)
 & \(8.63823\,10^{-16}\) & \(8.96440\,10^{-16}\)
 & \(3.26175\,10^{-17}\) & \(-18.8303\) & \(-3.6386\) \\
 \(1/2\) & \(1.51481\,10^{-3}\) & \(3.09348\,10^{-18}\)
 & \(3.38297\,10^{-18}\) & \(3.41924\,10^{-18}\)
 & \(3.62775\,10^{-20}\) & \(-9.5274\) & \(-1.0610\) \\
 \(1/4\) & \(7.63878\,10^{-4}\) & \(1.26108\,10^{-20}\)
 & \(1.32008\,10^{-20}\) & \(1.32405\,10^{-20}\)
 & \(3.96463\,10^{-23}\) & \(-4.7559\) & \(-0.2994\) \\
 \(1/8\) & \(3.83577\,10^{-4}\) & \(5.03276\,10^{-23}\)
 & \(5.15050\,10^{-23}\) & \(5.15478\,10^{-23}\)
 & \(4.27630\,10^{-26}\) & \(-2.3671\) & \(-0.0830\) \\
 \bottomrule
 \end{tabular}
\end{table}

\begin{figure}[!htbp]
 \centering
 \includegraphics[width=0.84\linewidth]{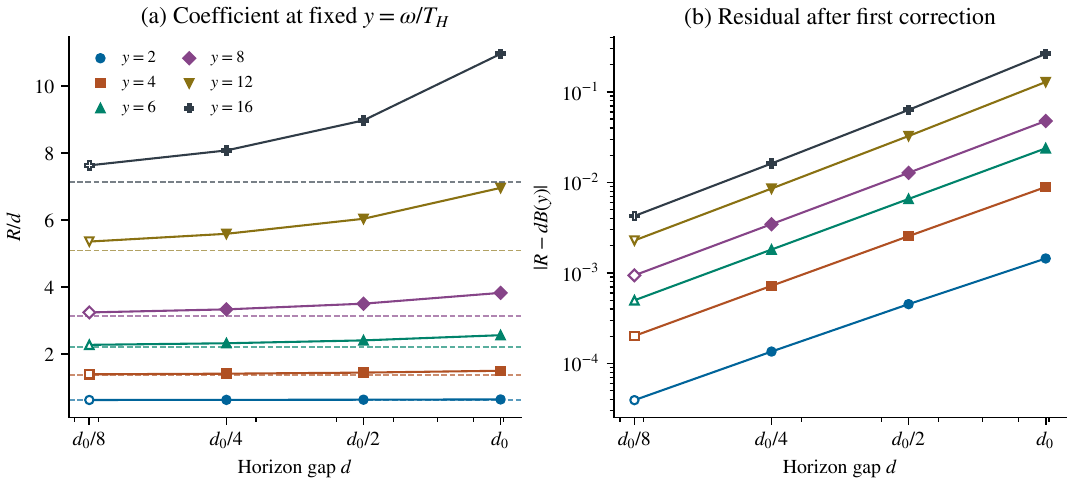}
 \caption{Fixed-frequency convergence at \(y=2,4,6,8,12,16\), with \(R=\Gamma_{\mathrm{num}}/\gref-1\). (a) The scaled difference \(R/d\) and the analytic \(B_{\mathrm H}(y)\) (dashed). (b) The residual \(\lvert R-d B_{\mathrm H}(y)\rvert\). Open markers indicate the held-out gap \(d_0/8\).}
 \label{fig:residuals}
\end{figure}

Figure~\ref{fig:residuals} compares \(R/d\) with \(B_{\mathrm H}(y)\) and shows how the residual after subtracting \(dB_{\mathrm H}(y)\) decreases with the gap. The prediction at \(d_0/8\) was fixed before its numerical evaluation.

\FloatBarrier
\clearpage
\twocolumngrid
\bibliographystyle{apsrev4-2}
\bibliography{references}

@article{Hawking1975,
  author = {Hawking, S. W.},
  title = {Particle creation by black holes},
  journal = {Commun. Math. Phys.},
  volume = {43},
  pages = {199--220},
  year = {1975},
  doi = {10.1007/BF02345020}
}

@article{Page1976,
  author = {Page, Don N.},
  title = {Particle emission rates from a black hole: Massless particles from an uncharged, nonrotating hole},
  journal = {Phys. Rev. D},
  volume = {13},
  pages = {198--206},
  year = {1976},
  doi = {10.1103/PhysRevD.13.198}
}

@article{Hayward2006,
  author = {Hayward, Sean A.},
  title = {Formation and evaporation of nonsingular black holes},
  journal = {Phys. Rev. Lett.},
  volume = {96},
  pages = {031103},
  year = {2006},
  doi = {10.1103/PhysRevLett.96.031103},
  note = {\href{https://arxiv.org/abs/gr-qc/0506126v2}{arXiv:gr-qc/0506126v2}}
}

@article{CalzaPedrottiVagnozzi2025,
  author = {Calz{\`a}, Marco and Pedrotti, Davide and Vagnozzi, Sunny},
  title = {Primordial regular black holes as all the dark matter. {I}. Time-radial-symmetric metrics},
  journal = {Phys. Rev. D},
  volume = {111},
  pages = {024009},
  year = {2025},
  doi = {10.1103/PhysRevD.111.024009},
  note = {\href{https://arxiv.org/abs/2409.02804v3}{arXiv:2409.02804v3}}
}

@misc{CalzaSpina2026,
  author = {Calz{\`a}, Marco and Spina, Andrea},
  title = {Gray-body factors: Method matters {II}, {WKB} vs {GrayHawk}},
  year = {2026},
  howpublished = {\href{https://arxiv.org/abs/2611.15889v1}{arXiv:2611.15889v1}}
}

@article{YangLim2022,
  author = {Yang, Ziou and Lim, Yen-Kheng},
  title = {Structure of test magnetic fields and charged particle motion around the {Hayward} spacetime},
  journal = {Phys. Rev. D},
  volume = {105},
  pages = {124045},
  year = {2022},
  doi = {10.1103/PhysRevD.105.124045},
  note = {\href{https://arxiv.org/abs/2203.12124v2}{arXiv:2203.12124v2}}
}

@article{Gubser1997,
  author = {Gubser, Steven S.},
  title = {Absorption of photons and fermions by black holes in four dimensions},
  journal = {Phys. Rev. D},
  volume = {56},
  pages = {7854--7868},
  year = {1997},
  doi = {10.1103/PhysRevD.56.7854},
  note = {\href{https://arxiv.org/abs/hep-th/9706100v2}{arXiv:hep-th/9706100v2}}
}

@article{CveticLarsen1998,
  author = {Cveti{\v c}, Mirjam and Larsen, Finn},
  title = {Greybody factors for black holes in four dimensions: Particles with spin},
  journal = {Phys. Rev. D},
  volume = {57},
  pages = {6297--6310},
  year = {1998},
  doi = {10.1103/PhysRevD.57.6297},
  note = {\href{https://arxiv.org/abs/hep-th/9712118v1}{arXiv:hep-th/9712118v1}}
}

@misc{DLMF,
  key = {{\relax DLMF}},
  title = {{NIST Digital Library of Mathematical Functions}},
  year = {2026},
  howpublished = {\url{https://dlmf.nist.gov/}, Release 1.2.8 of 2026-09-15},
  note = {edited by F.~W.~J. Olver \emph{et~al.}}
}

@article{ArbeyCalzaPerezGonzalez2025,
  author = {Arbey, Alexandre and Calz{\`a}, Marco and Perez-Gonzalez, Yuber F.},
  title = {Gray-body factors: Method matters},
  journal = {Phys. Dark Universe},
  volume = {48},
  pages = {101903},
  year = {2025},
  doi = {10.1016/j.dark.2025.101903},
  eprint = {2502.17240},
  archivePrefix = {arXiv}
}

@article{ArbeyEtAl2021,
  author = {Arbey, Alexandre and Auffinger, J{\'e}r{\'e}my and Geiller, Marc and Livine, Etera R. and Sartini, Francesco},
  title = {Hawking radiation by spherically-symmetric static black holes for all spins. {I}. {Teukolsky} equations and potentials},
  journal = {Phys. Rev. D},
  volume = {103},
  pages = {104010},
  year = {2021},
  doi = {10.1103/PhysRevD.103.104010},
  eprint = {2101.02951},
  archivePrefix = {arXiv}
}

@article{AyonBeatoGarcia1998,
  author = {Ay{\'o}n-Beato, Eloy and Garc{\'i}a, Alberto},
  title = {Regular Black Hole in General Relativity Coupled to Nonlinear Electrodynamics},
  journal = {Phys. Rev. Lett.},
  volume = {80},
  pages = {5056--5059},
  year = {1998},
  doi = {10.1103/PhysRevLett.80.5056},
  eprint = {gr-qc/9911046},
  archivePrefix = {arXiv}
}

@article{Calza2025GrayHawk,
  author = {Calz{\`a}, Marco},
  title = {{GrayHawk}: A public code for calculating the Gray Body Factors of massless fields around spherically symmetric Black Holes},
  journal = {Phys. Dark Universe},
  volume = {48},
  pages = {101900},
  year = {2025},
  doi = {10.1016/j.dark.2025.101900},
  eprint = {2502.04041},
  archivePrefix = {arXiv}
}

@article{CrispinoHiguchiMatsas2001,
  author = {Crispino, Lu{\'i}s C. B. and Higuchi, Atsushi and Matsas, George E. A.},
  title = {Quantization of the electromagnetic field outside static black holes and its application to low-energy phenomena},
  journal = {Phys. Rev. D},
  volume = {63},
  pages = {124008},
  year = {2001},
  doi = {10.1103/PhysRevD.63.124008},
  note = {Erratum in \href{https://doi.org/10.1103/PhysRevD.80.029906}{Phys. Rev. D \textbf{80}, 029906 (2009)}}
}

@article{CveticLarsen1997,
  author = {Cveti{\v c}, Mirjam and Larsen, Finn},
  title = {Greybody factors for rotating black holes in four dimensions},
  journal = {Nucl. Phys. B},
  volume = {506},
  pages = {107--120},
  year = {1997},
  doi = {10.1016/S0550-3213(97)00541-5},
  eprint = {hep-th/9706071},
  archivePrefix = {arXiv}
}

@article{DasGibbonsMathur1997,
  author = {Das, Sumit R. and Gibbons, Gary W. and Mathur, Samir D.},
  title = {Universality of Low Energy Absorption Cross Sections for Black Holes},
  journal = {Phys. Rev. Lett.},
  volume = {78},
  pages = {417--419},
  year = {1997},
  doi = {10.1103/PhysRevLett.78.417},
  eprint = {hep-th/9609052},
  archivePrefix = {arXiv}
}

@article{Dymnikova1992,
  author = {Dymnikova, Irina},
  title = {Vacuum nonsingular black hole},
  journal = {Gen. Relativ. Gravit.},
  volume = {24},
  pages = {235--242},
  year = {1992},
  doi = {10.1007/BF00760226}
}

@article{FanWang2016,
  author = {Fan, Zhong-Ying and Wang, Xiaobao},
  title = {Construction of regular black holes in general relativity},
  journal = {Phys. Rev. D},
  volume = {94},
  pages = {124027},
  year = {2016},
  doi = {10.1103/PhysRevD.94.124027},
  eprint = {1610.02636},
  archivePrefix = {arXiv}
}

@article{HarmarkNatarioSchiappa2010,
  author = {Harmark, Troels and Nat{\'a}rio, Jos{\'e} and Schiappa, Ricardo},
  title = {Greybody factors for {\(d\)}-dimensional black holes},
  journal = {Adv. Theor. Math. Phys.},
  volume = {14},
  pages = {727--793},
  year = {2010},
  eprint = {0708.0017},
  archivePrefix = {arXiv}
}

@article{Konoplya2023,
  author = {Konoplya, R. A.},
  title = {Quasinormal modes and grey-body factors of regular black holes with a scalar hair from the Effective Field Theory},
  journal = {J. Cosmol. Astropart. Phys.},
  volume = {2023},
  number = {07},
  pages = {001},
  year = {2023},
  doi = {10.1088/1475-7516/2023/07/001},
  eprint = {2305.09187},
  archivePrefix = {arXiv}
}

@article{KonoplyaOvchinnikovAhmedov2023,
  author = {Konoplya, R. A. and Ovchinnikov, D. and Ahmedov, B.},
  title = {Bardeen spacetime as a quantum corrected Schwarzschild black hole: Quasinormal modes and Hawking radiation},
  journal = {Phys. Rev. D},
  volume = {108},
  pages = {104054},
  year = {2023},
  doi = {10.1103/PhysRevD.108.104054},
  eprint = {2307.10801},
  archivePrefix = {arXiv}
}

@article{KonoplyaZhidenkoZinhailo2019,
  author = {Konoplya, R. A. and Zhidenko, A. and Zinhailo, A. F.},
  title = {Higher order {WKB} formula for quasinormal modes and grey-body factors: recipes for quick and accurate calculations},
  journal = {Class. Quantum Grav.},
  volume = {36},
  pages = {155002},
  year = {2019},
  doi = {10.1088/1361-6382/ab2e25},
  eprint = {1904.10333},
  archivePrefix = {arXiv}
}

@article{MaldacenaStrominger1997,
  author = {Maldacena, Juan M. and Strominger, Andrew},
  title = {Black Hole Greybody Factors and {D}-Brane Spectroscopy},
  journal = {Phys. Rev. D},
  volume = {55},
  pages = {861--870},
  year = {1997},
  doi = {10.1103/PhysRevD.55.861},
  eprint = {hep-th/9609026},
  archivePrefix = {arXiv}
}

@article{Unruh1976,
  author = {Unruh, W. G.},
  title = {Absorption cross section of small black holes},
  journal = {Phys. Rev. D},
  volume = {14},
  pages = {3251--3259},
  year = {1976},
  doi = {10.1103/PhysRevD.14.3251}
}
\end{document}